\documentstyle[times,pramana,epsf,floats,amssymb]{ias}

\def\plotone#1{\centering \leavevmode
\epsfxsize= 0.8\columnwidth \epsfbox{#1}}

\def\lsim{\mathrel{\rlap{\lower4pt\hbox{\hskip1pt$\sim$}}
    \raise1pt\hbox{$<$}}}                
\def\gsim{\mathrel{\rlap{\lower4pt\hbox{\hskip1pt$\sim$}}
    \raise1pt\hbox{$>$}}}                

\begin{document}
\mark{{Statistical anisotropy of CMB }{T. Souradeep and A. Hajian}}
\title{Statistical isotropy of the Cosmic Microwave Background}

\author{{\bf Tarun Souradeep} and Amir Hajian} \address{}  \pacs{2.0}
\abstract{The breakdown of statistical homogeneity and isotropy of
  cosmic perturbations is a generic feature of ultra large scale
  structure of the cosmos, in particular, of non trivial cosmic
  topology. The statistical isotropy (SI) of the Cosmic Microwave
  Background temperature fluctuations (CMB anisotropy) is sensitive to
  this breakdown on the largest scales comparable to, and even beyond
  the cosmic horizon.  We propose a set of measures, $\kappa_\ell$
  ($\ell=1,2,3, \ldots$) which for non-zero values indicate and
  quantify statistical isotropy violations in a CMB map.  We
  numerically compute the predicted $\kappa_\ell$ spectra for CMB
  anisotropy in flat torus universe models. Characteristic signature
  of different models in the $\kappa_\ell$ spectrum are noted. }

\maketitle

In standard cosmology, the Cosmic Microwave Background (CMB)
anisotropy is expected to be statistically isotropic, i.e.,
statistical expectation values of the temperature fluctuations $\Delta
T(\hat q)$ are preserved under rotations of the sky. In particular,
the angular correlation function $C(\hat{q},\,
\hat{q}^\prime)\equiv\langle\Delta T(\hat q)\Delta T(\hat
q^\prime)\rangle$ is rotationally invariant. In spherical harmonic
space, where $\Delta T(\hat q)= \sum_{lm}a_{lm} Y_{lm}(\hat q)$ this
translates to a diagonal $\langle a_{lm} a^*_{l^\prime
  m^\prime}\rangle=C_{l} \delta_{ll^\prime}\delta_{mm^\prime}$ where
$C_l$ is the widely used angular power spectrum.

In absence of statistical isotropy, $C(\hat{q},\hat{q}')$ is estimated
by a single product $\Delta T(\hat q) \Delta T(\hat q')$ and hence is
poorly determined from a single realization. Although it not possible
to estimate each element of the full correlation function
$C(\hat{q},\hat{q}')$, some measures of statistical anisotropy of the
CMB map can be estimated through suitably weighted angular averages of
$\Delta T(\hat q) \Delta T(\hat q')$. The angular averaging procedure
should be such that the measure involves averaging over sufficient
number of independent `measurements', but should ensure that the
averaging does not erase all the signature of statistical anisotropy.
Another important desirable property is that measure be independent of
the overall orientation of the sky. Based on these considerations, we
propose a set of measures $\kappa_\ell$ of statistical an isotropy
given by
\begin{equation}\label{kl}
 \kappa_\ell= \int\!\!  d\Omega\!\!\int\!\!  d\Omega^\prime 
 \left[\frac{(2\ell+1)}{8\pi^2}\!\!\int d{\mathcal R} \chi_\ell({\mathcal R})
C({\mathcal R}\hat{q},{\mathcal R}\hat{q}^\prime)\right]^2\!\!\!\!,
\end{equation}
where $ C({\mathcal R}\hat{q},\, {\mathcal R}\hat{q}^\prime)$ is the
two point correlation between ${\mathcal R}\hat{q}\,$ and $ {\mathcal
  R}\hat{q}^\prime$ obtained by rotating $\hat{q}$ and
$\hat{q}^\prime$ by an element ${\mathcal R}$ of the rotation
group~\cite{us_apj}.  The measures $\kappa_\ell$ involve angular
average of the correlation weighed by the characteristic function of
the rotation group $ \chi_\ell({\mathcal R})=\sum_{M}
D_{MM}^{\ell}({\mathcal R})$ where $ D_{MM^\prime}^{\ell}$ are the
Wigner D-functions~\cite{Var}.  When SI holds $ C({\mathcal
  R}\hat{q},\, {\mathcal R}\hat{q}^\prime)\,=\,C(\hat{q},\,
\hat{q}^\prime)$ is invariant under rotation, and eq.~(\ref{kl}) gives
$\kappa_\ell = \kappa_0\, \delta_{\ell 0}$ due to the orthonormality
of $\chi_\ell({\mathcal R})$.  {\em Hence, non-zero $\kappa_\ell$ for
  $\ell >0$ measure violation of statistical isotropy.}
 
The measure $\kappa_\ell$ has a clear interpretation in harmonic
space.  The two point correlation $C(\hat{q},\, \hat{q}^\prime)$ can
be expanded in terms of the orthonormal set of bipolar spherical
harmonics whose coefficients $A_{ll^\prime}^{\ell M}$ are related to
`angular momentum' sum over the covariances $\langle
a_{lm}a^*_{l^\prime m^\prime}\rangle$ as $A_{ll^\prime}^{\ell M} =
\sum_{m m^\prime} \langle a_{lm}a^*_{l^\prime m^\prime}\rangle \, \,
(-1)^{m^\prime} \mathfrak{ C}^{\ell M}_{lml^\prime -m^\prime}\,$ ,
where $\mathfrak{C}^{\ell M}_{lml^\prime m^\prime}$ are Clebsch-Gordan
coefficients~\cite{Var}.  The estimation of $\kappa_\ell$ from a CMB
map is discussed in~\cite{us_apj}.

The detection of statistical isotropy (SI) violations can have
exciting and far-reaching implication for cosmology.  The realization
that the universe with the same local geometry has many different
choices of global topology has been a theoretical curiosity as old as
modern cosmology. Motivations for cosmic topology and their
consequences have been extensively studied~\cite{costop}. CMB
anisotropy measurements have brought cosmic topology from the realm of
theoretical possibility to within the grasp of
observations~\cite{costop,bps}.  {\em A generic consequence of cosmic
  topology is the breaking of statistical isotropy in characteristic
  patterns determined by the photon geodesic structure of the
  manifold.}  Global isotropy of space is violated in all multi
connected models (except $S^3/Z_2$).  In cosmology, the Dirichlet
domain (DD) constructed around the observer represents the universe as
`seen' by the observer.  The SI breakdown is apparent in the principal
axes present in the shape of the DD constructed with the observer
located at the base-point~\cite{us_prl}.

In this paper we compute and study the $\kappa_\ell$ spectrum of SI
violation arising in flat (Euclidean) simple torus models with a
cubic, cuboidal and more generally, parallelepiped (squeezed)
fundamental domain. The CMB anisotropy in torus spaces has been well
studied~\cite{costop}.  We can relate the $\kappa_\ell$ spectrum to
the principal directions normal to pair of faces of the DD, their
relative orientation and the relative importance given by the distance
to the faces along them.  Along the most dominant axes, the distance
is minimum, and equals the {\em inradius}, $R_<$, the radii of largest
sphere fully enclosed within the DD~\cite{bps}.

The compact spaces with Euclidean geometry (zero curvature) have been
completely classified. In three dimensions, there are known to be six
possible topologies that lead to orientable spaces
~\cite{costop,wol94vin93}.  The simple flat torus, ${\cal M} = T^3$,
is obtained by identifying the universal cover ${\cal M}^u={\cal E}^3$
under a discrete group of translations along three non-degenerate
axes, ${\mathbf s_1,\mathbf s_2,\mathbf s_3}$: $ {\mathbf s_i} \to
{\mathbf s_i} + {\mathbf n} L_i$, where $L_i$ is the identification
length of the torus along $s_i$ and ${\bf n}$ is a vector with integer
components.  In the most general form, the fundamental domain (FD) is
a parallelepiped defined by three sides $L_i$ and the three angles
$\alpha_i$ between the axes (We call it squeezed torus).  If ${\mathbf
  s_i}$ are orthogonal then one gets cuboidal FD, which for equal
$L_i$ reduces to the cubic torus.  The cuboid and squeezed spaces
which can be obtained by a linear coordinate transformation ${\cal L}$
on cubic torus can have distinctly different global
symmetry~\footnote{For cubic torus the Dirichlet domain (DD) matches
  the fundamental domain (FD).  However, for torus spaces with cuboid
  and parallelepiped FD, the corresponding DD is very different, e.g.,
  hexagonal prism~\cite{wol94vin93,us_prl}.}.

We restrict attention to the case where CMB anisotropy arises entirely
at the sphere of last scattering (SLS) of radius $R_{*}$.  Invoking
method of images, the CMB correlation pattern on the SLS is known to
be dictated by the distribution of nearest `images' of the SLS on the
universal cover~\cite{bps}. The correlations are distorted even when
the SLS and its images do not intersect ($R_* < R_<$). When SLS
intersects its images the CMB sky is multiply imaged in characteristic
correlation pattern of pairs of circles~\cite{circles}.

\begin{figure}
\plotone{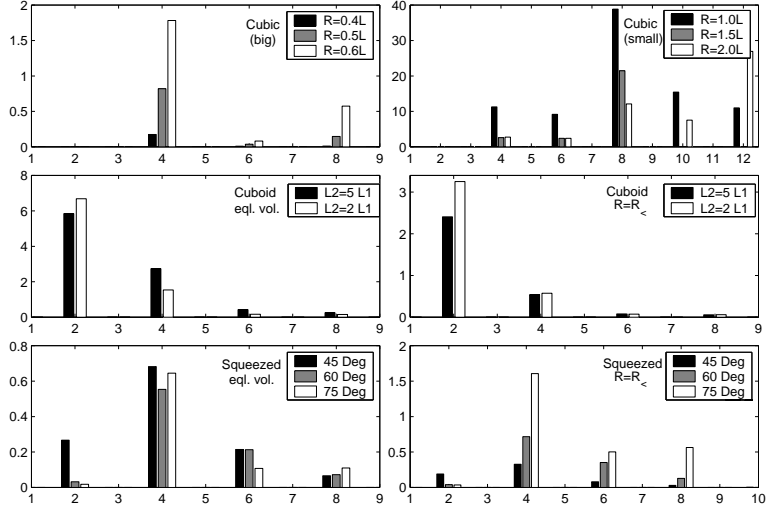}
  \caption{The $\kappa_{\ell}$ spectra for flat tori models are plotted.
    The top row panels are for cubic tori spaces. The left panel shows
    spaces of volume, $V_{\cal M}$, larger than the volume $V_{*}$
    contained in the sphere of last scattering (SLS) with $V_{\cal
      M}/V_{*} = 3.7,1.9,1.1$, respectively.  The right panel shows
    small spaces with $V_{\cal M}/V_{*} = 0.24,0.07,0.03$,
    respectively.  Note that $\kappa_2=0$ for cubic tori.  The middle
    panels consider cuboid tori with $1:5$ and $1:2$ ratio of
    identification lengths. The bottom panels show $\kappa_{\ell}$ for
    equal-sided squeezed tori with $\alpha=45^\circ,60^\circ$ and
    $75^\circ$. In the middle and bottom rows, the right panels show
    the case when radius of SLS, $R_*=R_<$ the inradius of the space.
    Here, the SLS just touches its nearest images which is at the
    threshold where CMB anisotropy is multiply imaged for larger
    $R_*$. The cases in the left panels of lower two rows have
    $V_{\cal M}/V_{*} =1$ and are at the divide between large and
    small spaces. }
  \label{kappaltorus}
\end{figure}

We compute $C({\hat q,\hat q^\prime})$ for CMB anisotropy in torus
space using regularized method of images~\cite{bps}.
Fig.~\ref{kappaltorus} plots the predicted $\kappa_\ell$ spectrum for
a number of cubic, cuboidal and squeezed torus spaces. We note the
following interesting results~:

{\bf i.} $\kappa_\ell=0$ for odd $\ell$ for all torus models. This
does not hold for compact space of non-zero curvature, e.g., compact
hyperbolic spaces.

{\bf ii.}  For cubic torus $\kappa_2=0$. $\kappa_2$ is non-zero for cuboidal 
and squeezed torus. This is a clear signature of non-cubic torus where the 
DD differs from the FD and has more than three principle axes.

{\bf iii.} For equal-sided squeezed torus, $\kappa_4$ , decreases as
$\alpha$ decreases from $90^\circ$ to $60^\circ$ as $R_<$ increases.
For $\alpha<60^\circ$ sharply increases with decreasing $\alpha$ as
$R_<$ decreases sharply.

{\bf iv.} $\kappa_2=0$ increases monotonically as $\alpha$ decreases
from $90^\circ$.

{\bf v.} The peak of $\kappa_\ell$ shifts to larger $\ell$ for small
spaces.

\noindent The results can be understood using  the
leading order terms of the correlation function in a torus where
$\kappa_\ell$ can be calculated analytically~\cite{us_prl}.

Preferred directions and statistically anisotropic CMB anisotropy have
been discussed in literature~\cite{fer_mag97bun_scot00}. When CMB
anisotropy is multiply imaged, the $\kappa_\ell$ spectrum corresponds
to a correlation pattern of matched pairs of circles~\cite{circles}.
The generic features of $\kappa_\ell$ spectrum are related to the
symmetries of correlation pattern. But $\kappa_\ell$ are sensitive to
SI violation even when CMB is not multiply imaged. Moreover
$\kappa_\ell$ have an advantage of being insensitive to the overall
orientation of the correlation features.  The $A^{\ell M}_{ll^\prime}$
signature, which was not discussed here, contains more details of the
SI violation. The estimation of $\kappa_\ell$ from a CMB map is
described~\cite{us_apj}.  Before ascribing the detected breakdown of
statistical anisotropy to cosmological or astrophysical effects, one
must carefully account for and model out other mundane sources of
statistical anisotropy in real data, such as, incomplete and
non-uniform sky coverage, beam anisotropy, foreground residuals and
statistically anisotropic noise.  These observational artifacts will
be discussed in future publications.


\begin{thebibliography}{refer 9999}
\bibitem{costop} G. F. R. Ellis, Gen. Rel. Grav. {\bf 2}, 7,(1971); M.
  Lachieze-Rey \& J. -P. Luminet, Phys. Rep.  {\bf 25}, 136, (1995);
  J. Levin, Phys. Rep. {\bf 365}, 251, (2002); G. Starkman, Class.
  Quantum Grav. {\bf 15}, 2529, (1998).  
\bibitem{wol94vin93} J. A. Wolf, {\it Space of Constant Curvature (5th
    ed.)}, (Publish or Perish, Inc., 1994); E. B. Vinberg, {\it
    Geometry II -- Spaces of constant curvature}, (Springer-Verlag,
  1993).
\bibitem{bps} J. R. Bond, D. Pogosyan \& T. Souradeep, Class. Quant.
  Grav. {\bf 15}, 2671, (1998); {\it ibid.}  Phys.  Rev. {\bf D
    62},043005, (2000);{\it ibid.},043006, (2000); T. Souradeep, in
  `The Universe', ed. N. Dadhich \& A. Kembhavi (Kluwer, 2000).
\bibitem{Var} D. A. Varshalovich, A. N. Moskalev,  V. K. Khersonskii, 
  {\it Quantum Theory of Angular Momentum} (World Scientific 1988). 
\bibitem{us_apj} A.  Hajian \& T. Souradeep, IUCAA preprint, astro-ph/0308001.
\bibitem{us_prl} Hajian, A.  \& Souradeep, T., {\it preprint} (astro-ph/0301590).
\bibitem{circles} N.J. Cornish, D.N. Spergel \& G. D. Starkman,  
Class. Quantum Grav., {\bf 15}, 2657, (1998).
\bibitem{fer_mag97bun_scot00} P. G. Ferreira \& J. Magueijo, Phys.
  Rev.  {\bf D56}, 4578, (1997); E. Bunn \& D. Scott, M.N.R.A.S., {\bf 313},
  331, (2000).
\end{thebibliography}
\end{document}